\documentclass[10pt]{article}
\usepackage{graphicx}
\usepackage{atlasunit}
\usepackage{subfigure}

\def\Title#1{\begin{center} {\Large #1 } \end{center}}
\def\Author#1{\begin{center}{ \sc #1} \end{center}}
\def\Address#1{\begin{center}{ \it #1} \end{center}}

\newcommand\pubblock{\rightline{\begin{tabular}{l} Proceedings of the Fifth Annual LHCP\\ \pubnumber\\
         \pubdate  \end{tabular}}}

\newenvironment{Abstract}{\begin{quotation} \begin{center} 
             \large ABSTRACT \end{center}\bigskip 
      \begin{center}\begin{large}}{\end{large}\end{center} \end{quotation}}

\newenvironment{Presented}{\begin{quotation} \begin{center} 
             PRESENTED AT\end{center}\bigskip 
      \begin{center}\begin{large}}{\end{large}\end{center} \end{quotation}}

\def\beq{\begin{equation}}
\def\eeq#1{\label{#1}\end{equation}}
\def\eeqn{\end{equation}}

\def\beqa{\begin{eqnarray}}
\def\eeqa#1{\label{#1}\end{eqnarray}}
\def\eeqan{\end{eqnarray}}

\let\bar=\overbar

\def\Dslash{\not{\hbox{\kern-4pt $D$}}}
\def\dslash{\not{\hbox{\kern-2pt $\del$}}}

\def\msb{{\bar{\ssstyle M \kern -1pt S}}}

\usepackage{color}

\newcommand\pubnumber{ ATL-PHYS-PROC-2017-205 }
\newcommand\pubdate{\today}

\def\affiliation{
On behalf of the ATLAS Experiment \\
Institute of high energy physics, Chinese Academy of Science,Beijing, 100049, China}

\begin{document}

% large size for the first page
\large
\begin{titlepage}
\pubblock

%% Change the title, name, abstract
%% Title 
\vfill
\Title{Search for the Standard Model Higgs boson
decaying to b-quark pairs with the ATLAS detector at the LHC }
\vfill

%  if you need to add the support use this, fill the \support definition above. 
%   \Author{ FIRSTNAME LASTNAME \support }
\Author{ Zhijun Liang }
\Address{\affiliation}
\vfill
\begin{Abstract}

After many years of searches, the Higgs boson was observed by the ATLAS and CMS collaborations
in July 2012. Since then, its properties have been measured using primarily the
bosonic decay channels: $H \to \gamma\gamma $, $H \to Z Z$, $H \to W^+ W^-$.
In order to probe the predictions of the Standard Model and the nature of the Higgs boson, it is also fundamental
to measure its couplings to fermions and, in particular, to quarks. In this paper we present the
ATLAS results in the search for the Higgs boson decaying to b-quark pairs using two different
production channels: associated production with a vector boson and vector boson fusion with an
additional hard photon, using 36.1 $\mathrm{fb^{-1}}$ the $pp$ collisions delivered by the LHC at a
center of mass energy of 13 $\mathrm{TeV}$.
\end{Abstract}
\vfill

% DO NOT CHANGE 
\begin{Presented}
The Fifth Annual Conference\\
 on Large Hadron Collider Physics \\
Shanghai Jiao Tong University, Shanghai, China\\ 
May 15-20, 2017
\end{Presented}
\vfill
\end{titlepage}
\def\thefootnote{\fnsymbol{footnote}}
\setcounter{footnote}{0}
%

% normal size for the rest
\normalsize 

%% Your paper should be entered below. 

\section{Introduction}

In 2012, a new particle with a mass of 125 $\mathrm{GeV}$ and its properties compatible with those expected for the Standard Model Higgs boson was discovered by the ATLAS and CMS Collaborations~\cite{Aad:2012tfa,Chatrchyan:2012ufa}.%,\cite{}.

Precise measurements of the Higgs boson couplings to other particles can provide further insight into the nature of electroweak symmetry breaking.
Even though the dominant decay of the Standard Model Higgs boson is to b quark pairs, the measurement of Higgs production in this particular signature has been proven to be a formidable challenge.
 The combined ATLAS$+$CMS results in Run 1 quote for a $2.6\sigma$ signal significance for $H\rightarrow b\bar{b}$~\cite{run1combine}. The ATLAS experiment~\cite{atlas} has searched for the Higgs decays to b-quark pairs using proton-proton collision data collected during the LHC Run 2, with a center of mass energy of 13 TeV. Given the huge background from b-quark pairs produced directly in the pp collisions, the search was done using two different sub-dominant production modes: associated production with a $W$ or a $Z$ boson (called together VH production) and vector boson fusion (VBF) with a high $\mathrm{p_T}$ photon accompanying the Higgs. Despite the reduction in the signal cross section due to the presence of the hard photon in the VBF channel, the dominant background $b\bar{b} + \gamma +$jet jet is suppressed significantly, leading to an improved signal over background ratio. In what follows, these two searches will be explained in detail.

\section{Search for $H \to b \bar{b}$ in associated production with a vector boson}
The search for the Higgs boson in this channel has been done considering the leptonic decays of the $W$ and $Z$ boson, which result in three different final states: $WH \to l \nu b \bar{b} $, $ZH \to l^+l^-b\bar{b} $ and $ZH \to \nu \nu b \bar{b}$. These channels are characterized by one, two or zero charged leptons in the final state, respectively. 
The experimental signatures are high transverse momentum ($\mathrm{p_T}$) isolated charged leptons or large missing transverse energy, in addition to the two b-jets.
The data used in this analysis were collected at a centre-of-mass energy of 13 $\mathrm{TeV}$ during the 2015 and 2016 running periods, and correspond to integrated luminosities of 3.2 $\pm$ 0.1 $\mathrm{fb^{-1}}$ and 32.9 $\pm$ 1.1 $\mathrm{fb^{-1}}$, respectively. They were collected using missing transverse momentum ($E_T^{miss}$) triggers for the 0-lepton and 1-lepton channels and single-lepton triggers for the 1- and 2-lepton channels. The selection benefited from better b-tagging performance with respect to the Run 1 analysis, due to an extra inner detector pixel layer which has been installed during the long shutdown of the LHC in 2014-15. This results in a light jet rejection factor of about 380 and a charm rejection factor around 12 for a b-tagging efficiency of 70\%.
Different analysis categories were defined according to the number of leptons in the final state, the number of jets in the final state (two or three for the 0-lepton and 1-lepton analyses, and two or more than two for the 2-lepton analysis), and the transverse momentum of the vector boson.
In the 0-lepton and 1-lepton channels, the vector boson was required to have a $p_T (H)> 150$ $\mathrm{GeV}$. In the 2-lepton channel two regions are considered,  75 $\mathrm{GeV} <p_T (H)<$ 150 $\mathrm{GeV}$  and $p_T (H)>$ 150 $\mathrm{GeV}$.
Events are further split into two categories according to jet multiplicity. In the 0-lepton and 1-lepton channels, events are considered with exactly two or exactly three jets. Events with four or more jets are rejected in these channels to reduce the large background arising from $t\bar{t}$ production. In the 2-lepton channel, extra sensitivity is gained by accepting events with higher jet multiplicity due to the lower level of the $t\bar{t}$ background, thus the categories become either exactly two jets or three or more jets. For simplicity, these two selection categories are referred to as the 2- and 3-jet categories for all three lepton channels.
A detailed description of the selection and the results in this channel can be found in Ref.~\cite{ATLAShbb2017}.

After the selection, the dominant backgrounds are $Z+$jets in the 0-lepton and 2-lepton channels, and top quark production and $W+$ jets in the 1-lepton channel. Further background reduction is obtained using a boosted decision tree (BDT) that combines different discriminating variables and is trained independently for the eight categories. The largest discrimination is obtained from the correlation of the invariant mass of the b-jet pairs and the separation of the two b-jets.

\begin{figure}[htb]
\centering
\includegraphics[height=2in]{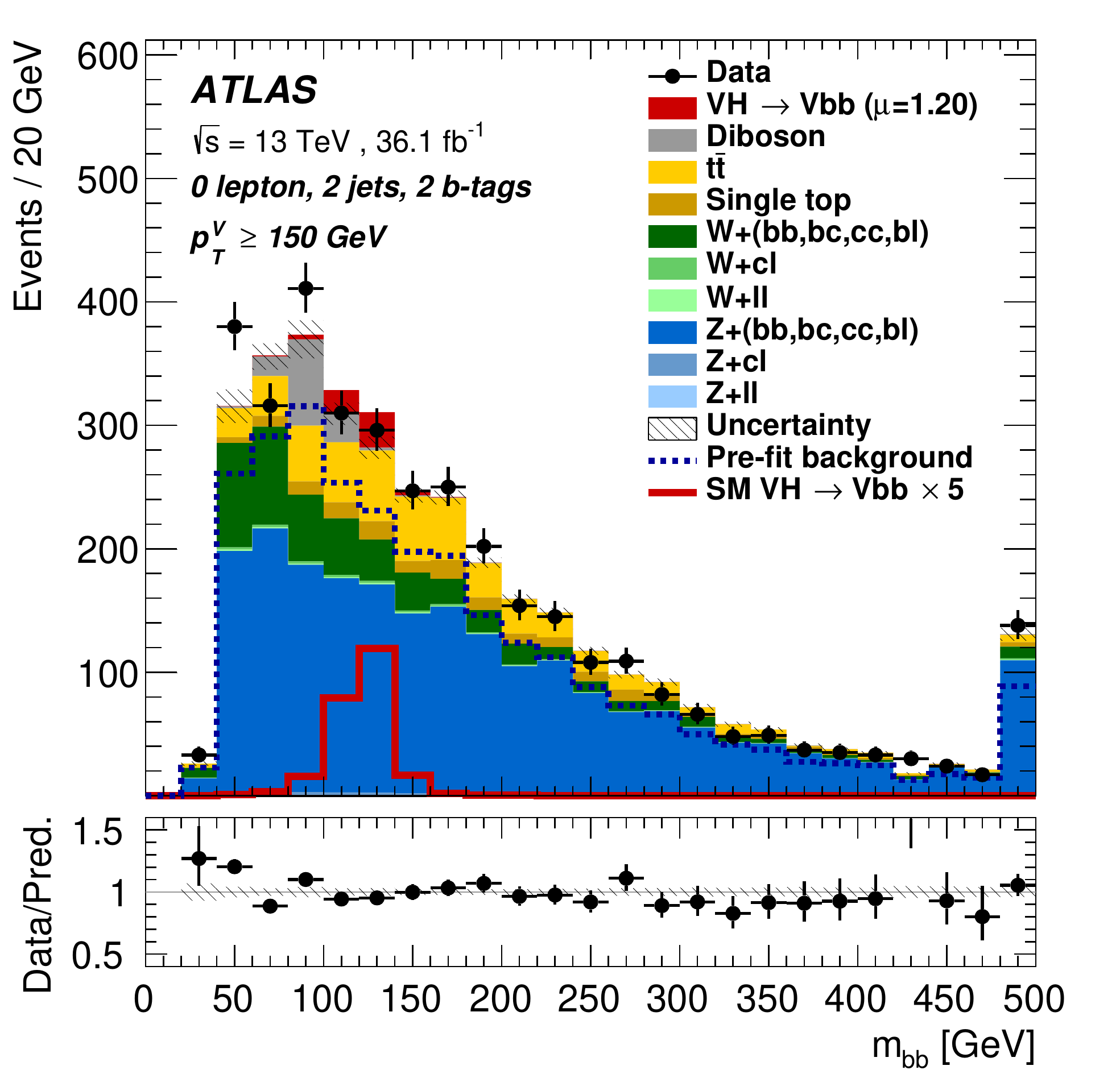}
\includegraphics[height=2in]{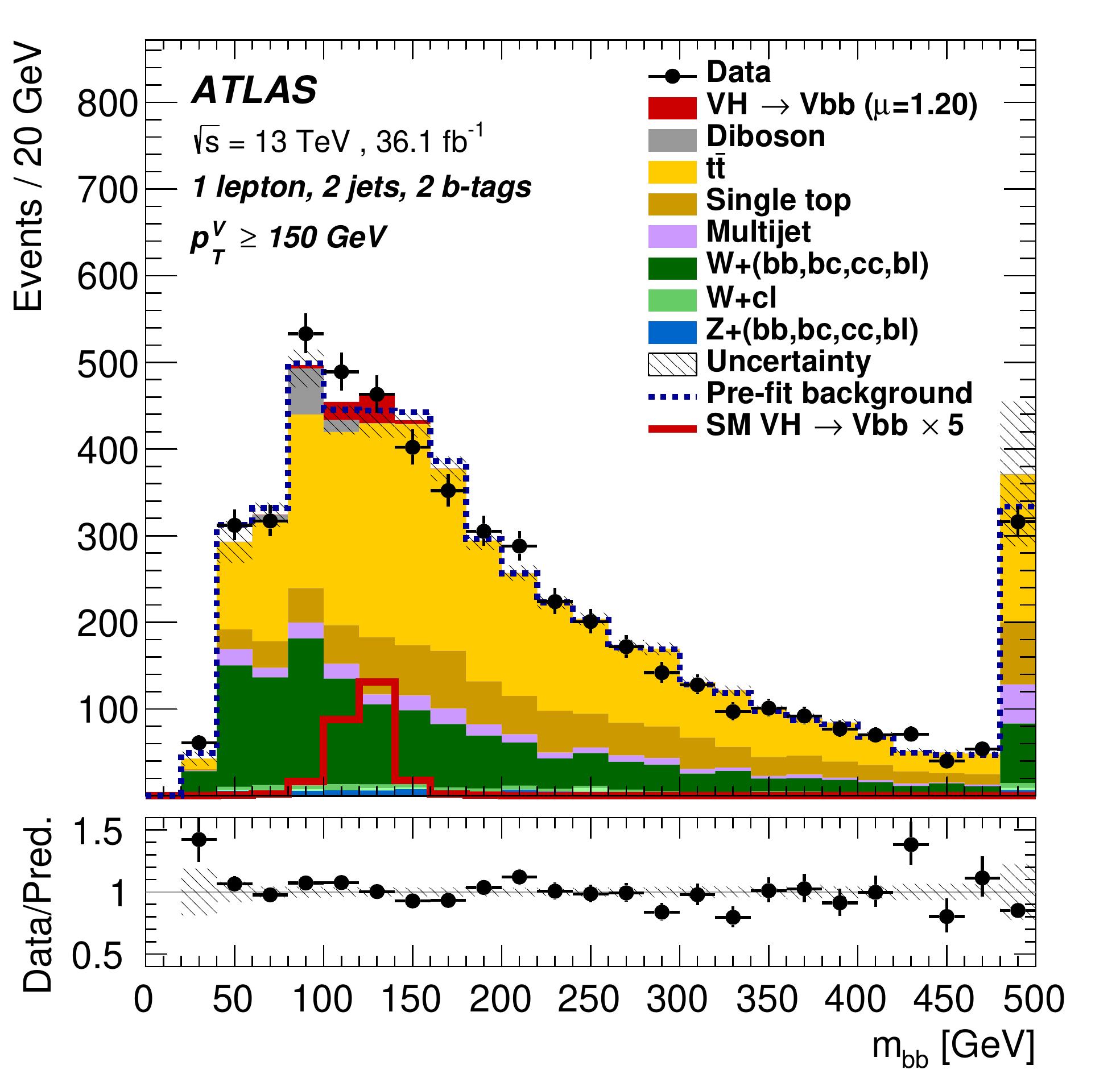}
\includegraphics[height=2in]{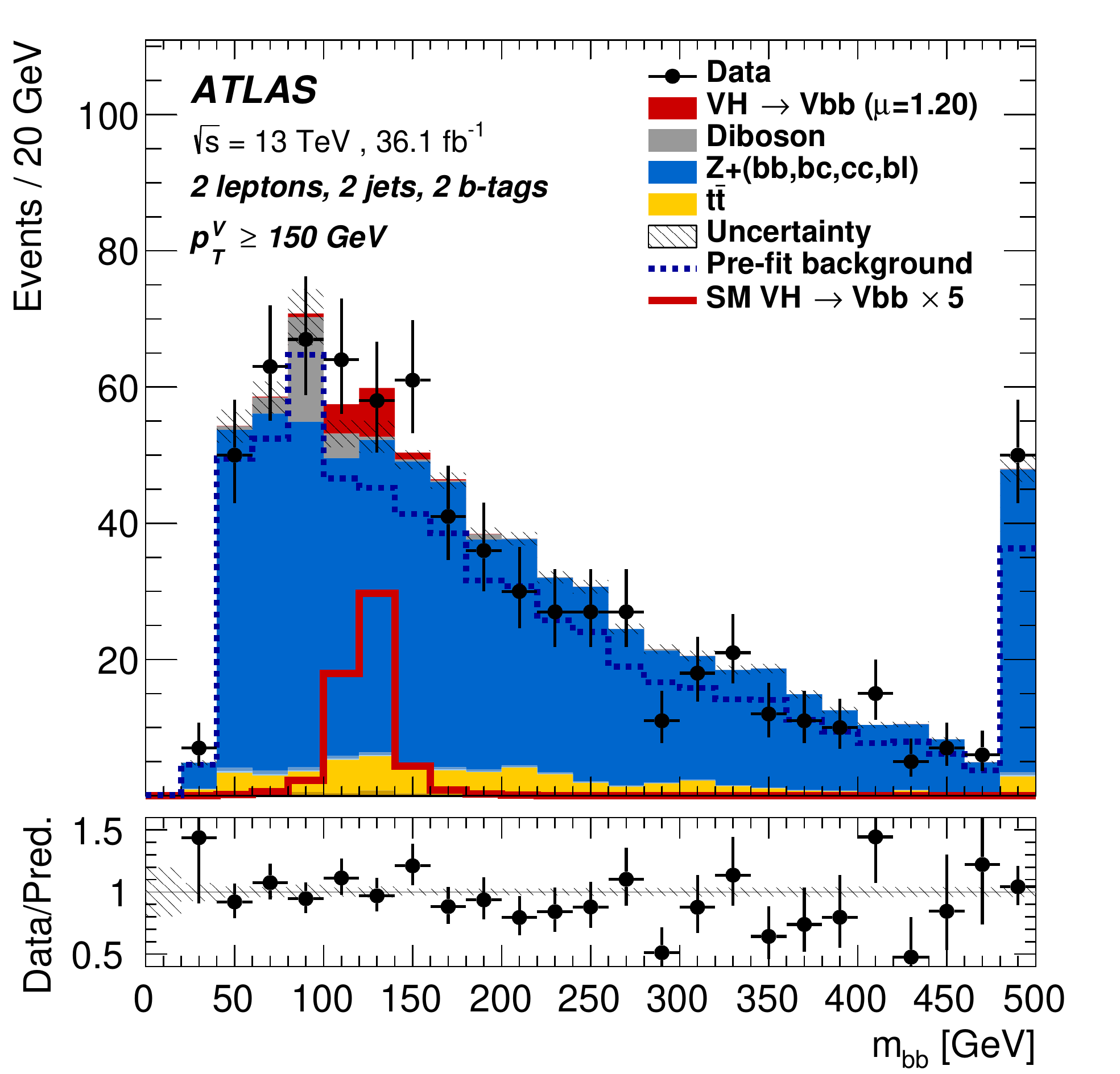}
\caption{ The post-fit distributions for $m_{bb}$ distribution in the 0-lepton (left), 1-lepton (middle) and 2-lepton (right) for 2 jet, 2 b-tag events in high $p_T^V$ region.The background contributions after the global likelihood fit are shown as filled histograms. The Higgs boson signal ($m_H =$  125 $\mathrm{GeV}$) is shown as a filled histogram on top of the fitted backgrounds normalised to the signal yield extracted from data (μ = 1.20), and unstacked as an unfilled histogram, scaled by the factor indicated in the legend. The entries in overflow are included in the last bin. The dashed histogram shows the total background as expected from the pre-fit MC simulation. The size of the combined statistical and systematic uncertainty for the sum of the fitted signal and background is indicated by the hatched band. The ratio of the data to the sum of the fitted signal and background is shown in the lower panel~\cite{ATLAShbb2017}.}
\label{fig:figure1}
\end{figure}

%Two new variables that were not used in previous analyses were included, corresponding to the mass of the top quark in $t\bar{t}$ events and to the difference in pseudorapidity between the vector boson and the Higgs. They provide an improvement of 7\% in the expected signal significance.
The signal strength $\mu$, defined as the ratio of the observed cross section times branching ratio with respect to the SM expectation, is obtained from a profiled likelihood fit that takes into account all the analysis categories and uses the BDT discriminant as input. The dominant backgrounds are normalized in the fit. The distributions of the post-fit spectrum for $m_{bb}$ distribution in the 0-lepton, 1-lepton and 2-lepton for 2 jet, 2 b-tag events in high $p_T^V$ region for data and MC are shown in Fig.~\ref{fig:figure1}. The background normalizations are those resulting from the fit. The pre-fit background expectations are also shown. After the fit, there is good agreement between the data and the background expectations in all the three channels, taking into account the uncertainties.

The dominant experimental uncertainties originate from the flavour-tagging simulation-to-data efficiency correction factors, the jet energy scale corrections and the modelling of the jet energy resolution. Flavour-tagging simulation-to-data efficiency correction factors are derived separately for b-jets, c-jets and light-flavour jets. The approximate size of the uncertainty in the tagging efficiency is 2\% for b-jets, 10\% for c-jets and 30\% for light jets.

The dominant uncertainties on signal modelling is the uncertainties due to the parton-shower and underlying-event models. They are estimated by considering the difference between Powheg$+$MiNLO$+$Pythia8 and Powheg$+$MiNLO$+$Herwig7, as well as changes in the Pythia8 parton-shower tune. The approximate size of the uncertainty is about 15\%.

%The event yields as a function of the logarithm of the signal over background are presented in figure 2 (right), combining all the different analysis categories. The Higgs signal is expected to accumulate on the right-most bins. The data agrees well with the Standard Model background predictions, as shown on the data over background ratio in the lower panel. No clear excess of events with respect to the background expectations is seen.

\begin{figure}[htb]
\centering
\includegraphics[height=2in]{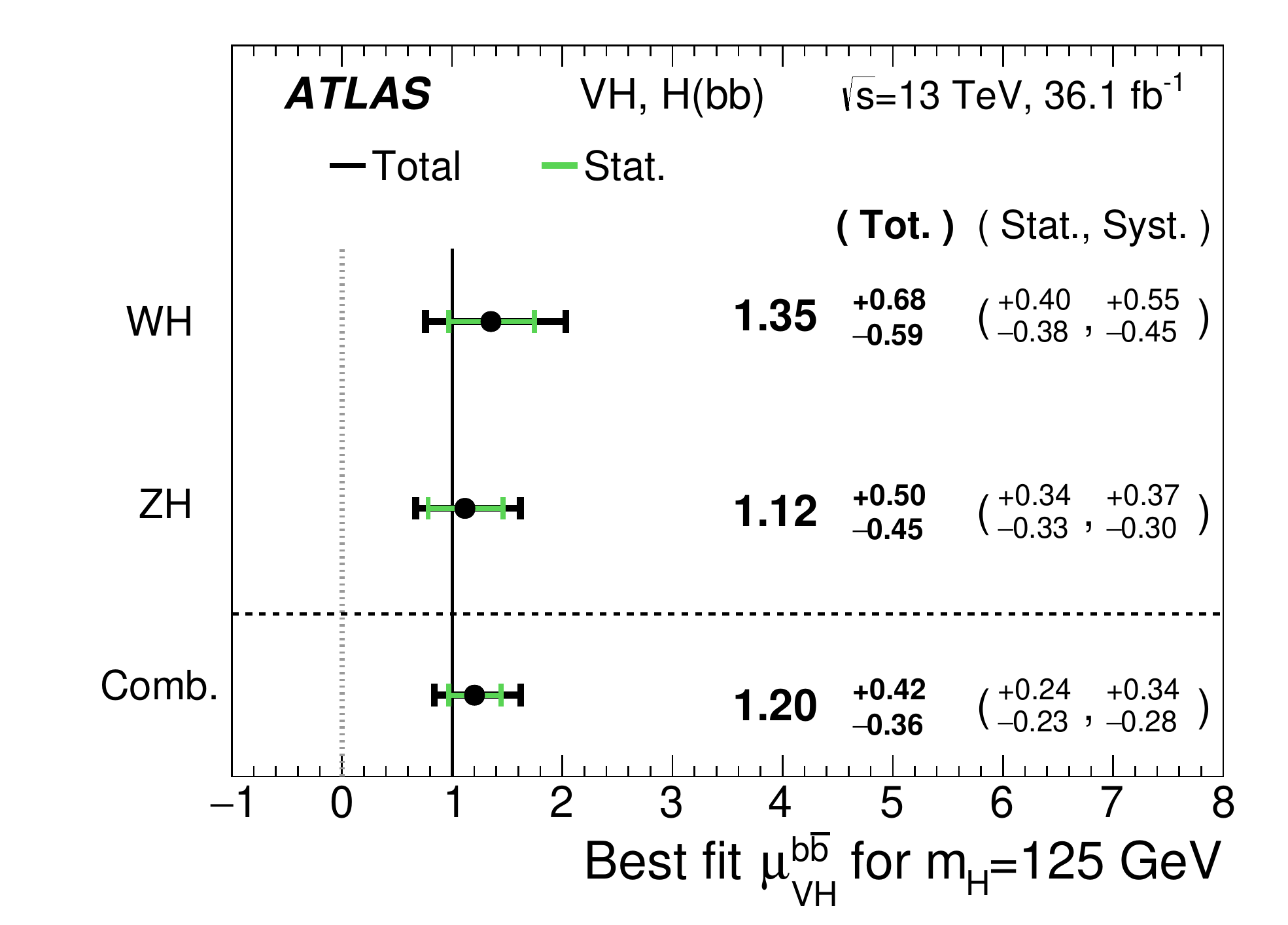}
\includegraphics[height=2in]{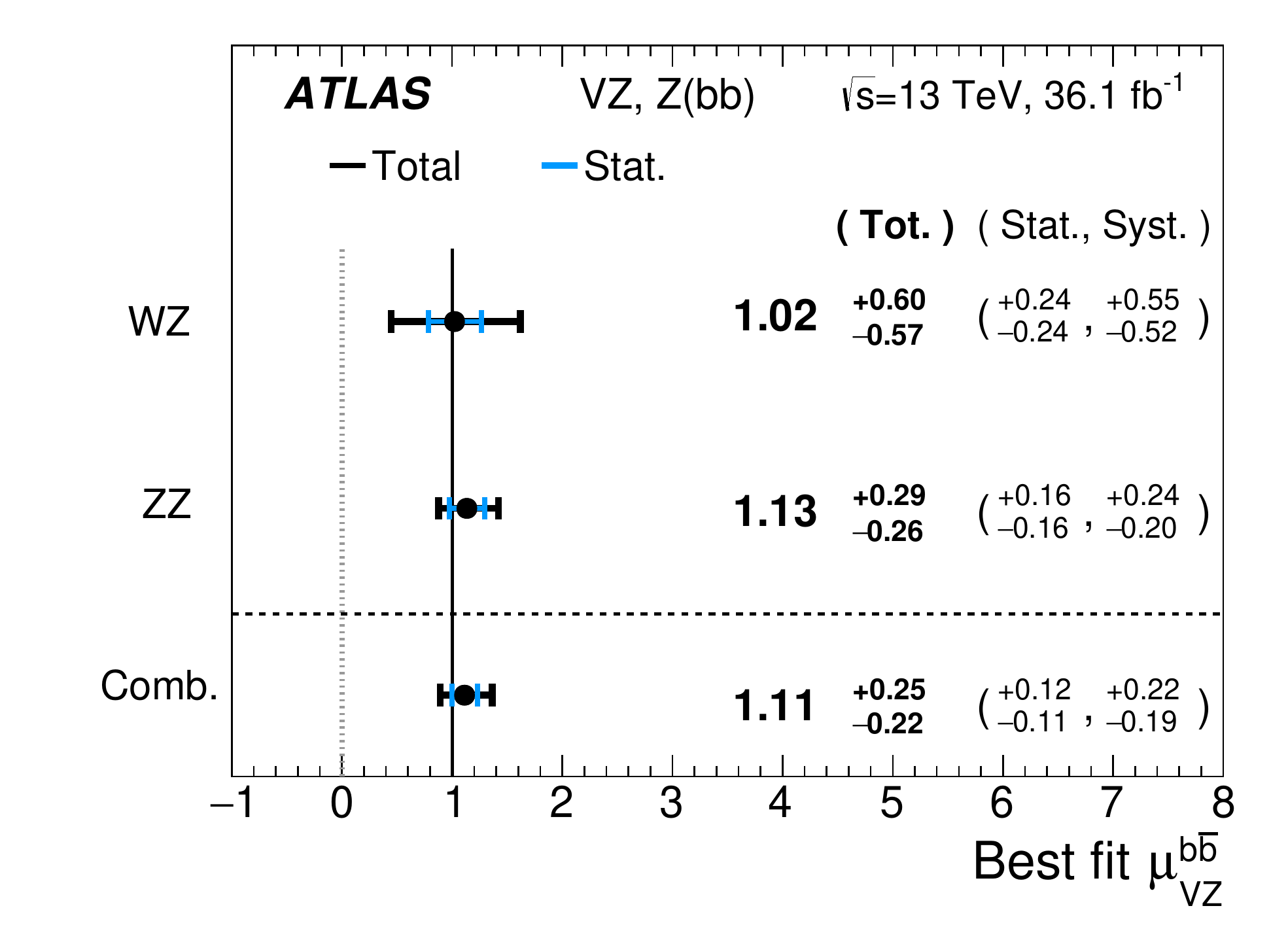}
\caption{  The fitted values of the signal strength parameter $\mu$ for the $WZ$ and $ZZ$ processes and their combination (left plot).  The fitted values of the Higgs boson signal strength parameter $\mu$ for $m_H$ = 125 $\mathrm{GeV}$ for the $WH$ and $ZH$ processes and their combination~\cite{ATLAShbb2017} (right plot).}
\label{fig:mu}
\end{figure}

The best fit signal strength is shown in Fig.~\ref{fig:mu}, for a 125 $\mathrm{GeV}$ Higgs boson mass in $WH$ and $ZH$ and for their combination.
The combined fitted value of the signal strength parameter for all the channels is $\mu_{VH}$ = $1.2 ^{+0.24}_{-0.23}$(stat.)$^{+0.34}_{-0.28}$(syst.). The observation corresponds to an excess with a significance of 3.5 standard deviations, to be compared to an expectation of 3.0 standard deviations. The $WH$ and $ZH$ production modes are observed with a significance of 2.4 and 2.6 standard deviations, respectively. The linear correlation term between the signal strengths related to the $WH$ and the $ZH$ production modes is 0.6\%. 
%For all channels combined the fitted value of the signal strength parameter is $\mu_{VZ}$ = $1.2 ^{+0.24}_{-0.23}$(stat.)$^{+0.34}_{-0.28}$(syst.). The observation corresponds to an excess with a significance of 3.5 standard deviations, to be compared to an expectation of 3.0 standard deviations. The $WH$ and $ZH$ production modes are observed with a significance of 2.4 and 2.6 standard deviations, respectively. The linear correlation term between the signal strengths related to the $WH$ and the $ZH$ production modes is 0.6\%. 

The fit was cross checked by searching for the standard model di-boson signal, $WZ + ZZ$,
where the Z boson decays to b-quark pairs. In this case, the observed signal significance is 5.8$\sigma$
with a signal strength of $\mu_{VZ}$ = $1.11 ^{+0.12}_{-0.11}$(stat.)$^{+0.22}_{-0.19}$(syst.), showing the validity of the procedure.
Analogously to the VH signal, fits are also performed with separate signal strength parameters for the $WZ$ and $ZZ$ production modes, and the results are shown in Fig.~\ref{fig:mu} .

\section{Search for $H \to b \bar{b}$ in vector boson fusion plus photon production}
One interesting possibility is to exploit the $H \to b \bar{b}$ process through vector (weak) boson fusion (VBF), which has been measured by ATLAS and CMS collaborations~\cite{ATLASvbfhbbrun1,CMSvbfhbb}. The sensitivities of these analyses are limited by very large contributions from non-resonant $b\bar{b}jj$ production and by difficulties in triggering on low-transverse momentum ($p_T$) b-jets.

\begin{figure}[htb]
\centering
\includegraphics[height=2in]{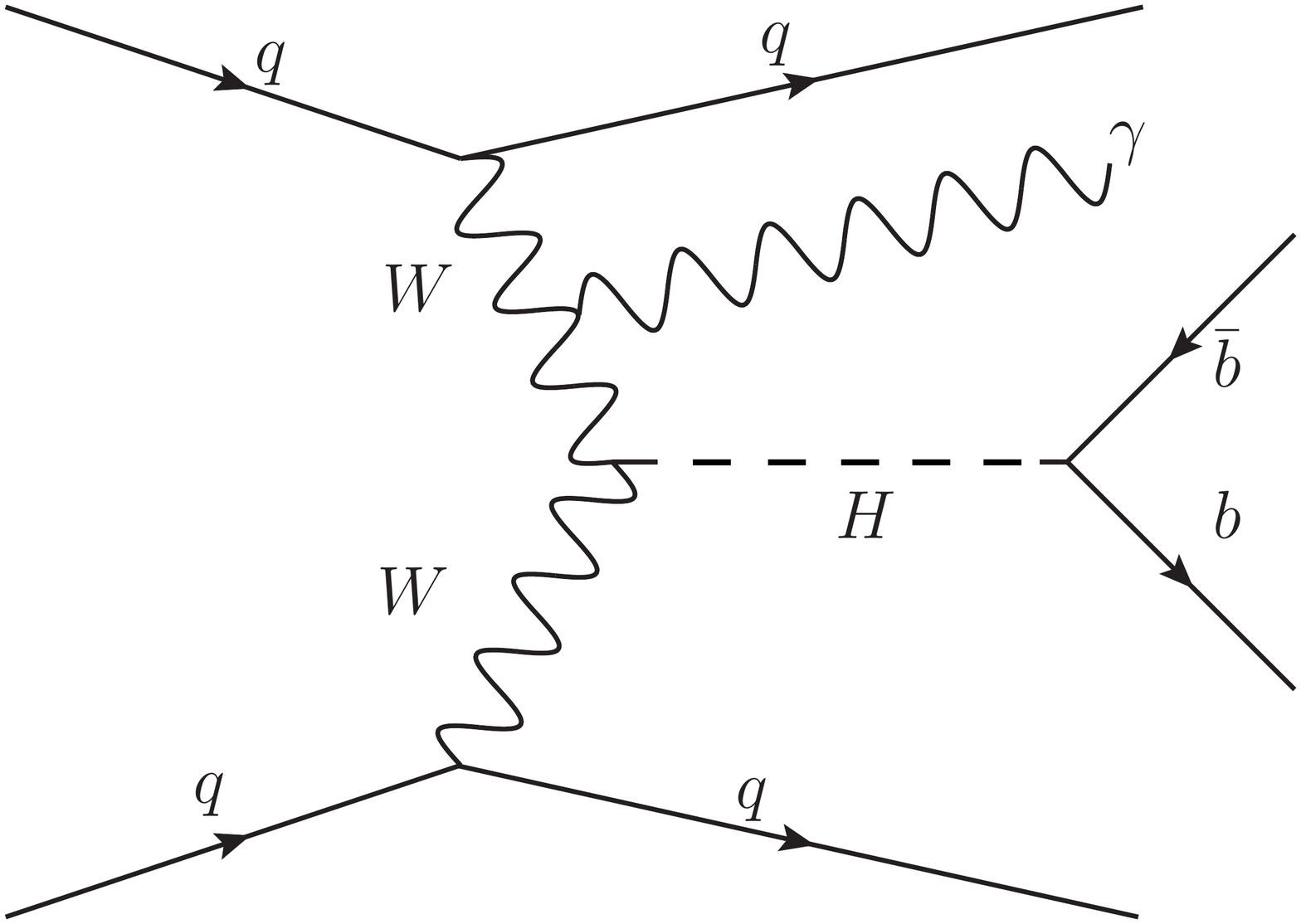}
\includegraphics[height=2in]{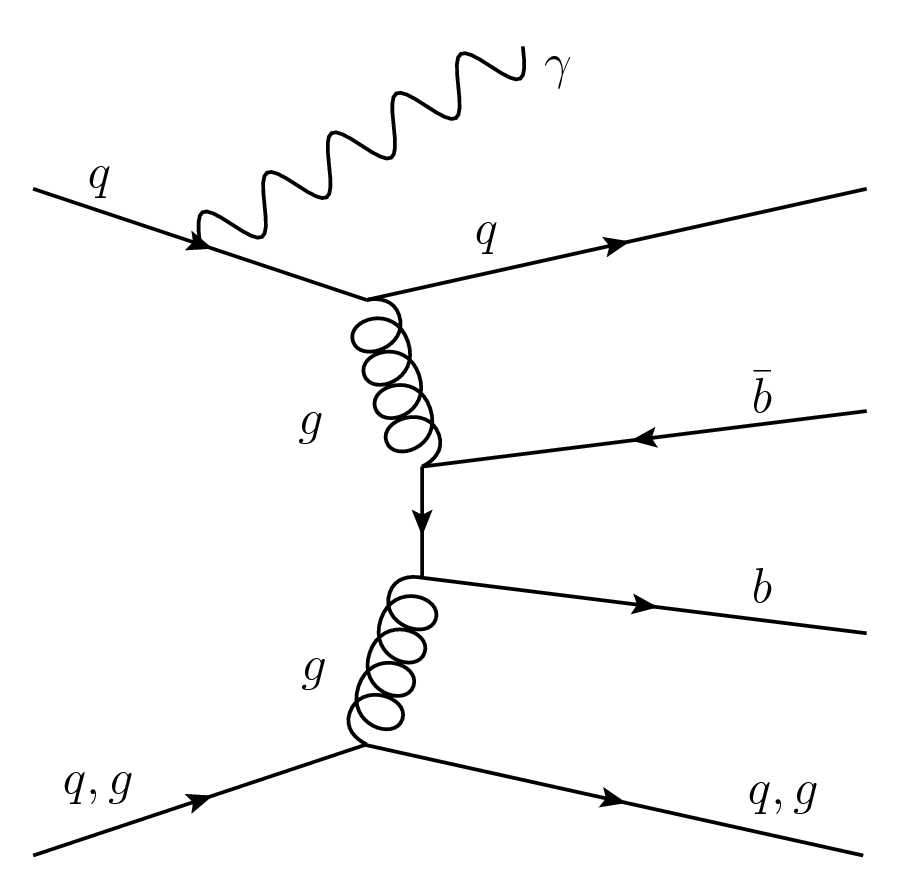}
\caption{ Representative leading-order Feynman diagrams for Higgs boson production via vector boson fusion in association with a photon(left). Representative Feynman diagrams for $b\bar{b}jj\gamma$ background process (right)~\cite{ATLASVBFhbbg}.  }
\label{fig:diagram}
\end{figure}

Previous study~\cite{theorypaper} has shown that requiring a high-$p_T$ photon in the final state can dramatically enhance the signal-to-background ratio in the VBF production mode. The photon may be radiated from an internal $W^{\pm}$ boson or from an incoming or outgoing quark, as shown in Fig.~\ref{fig:diagram}.

This channel is characterized by a high-$p_T$ photon, two b-jets from the Higgs decay and two high-$p_T$ forward jets with large invariant mass and large rapidity separation, that is the typical signature of VBF production. Events are pre-selected with a dedicated trigger that requires one photon and four jets. In offline selection, the event is required to have one good isolated photon, two b-tagged jets with an invariant mass $m_{bb}$ larger than 80 $\mathrm{GeV}$ and two extra jets with invariant mass $m_{jj} >$ 800 $\mathrm{GeV}$. The dominant background is multi-jet production of non-resonant heavy flavour jets in association with a photon. Other background processes include  $W$ or $Z$ production with a photon and additional jets, top-quark production with photons and jets.
To further improve the background rejection, a BDT discriminant is built using seven different variables that are not correlated with $m_{bb}$, such as the distance between the jet and the photon, the invariant mass of the VBF jets, the centrality of the photon and scalar $\mathrm{p_T}$ sum over the track-jets ($\mathrm{H_T}^{\mathrm{soft}}$). The centrality of the photon is defined as Eqn.~\ref{eqn:centrality}. $y_{j_1}$ and $y_{j_2}$ are the rapidity of the leading and subleading VBF jet respectively.
  \begin{equation} \mathrm{centrality}(\gamma)=\left| \frac{y_{\gamma}-\frac{y_{j_1}+y_{j_2}}{2}}{y_{j_1}-y_{j_2}} \right| ,
    \label{eqn:centrality}
  \end{equation}

\begin{figure}[htb]
\centering
\includegraphics[height=2in]{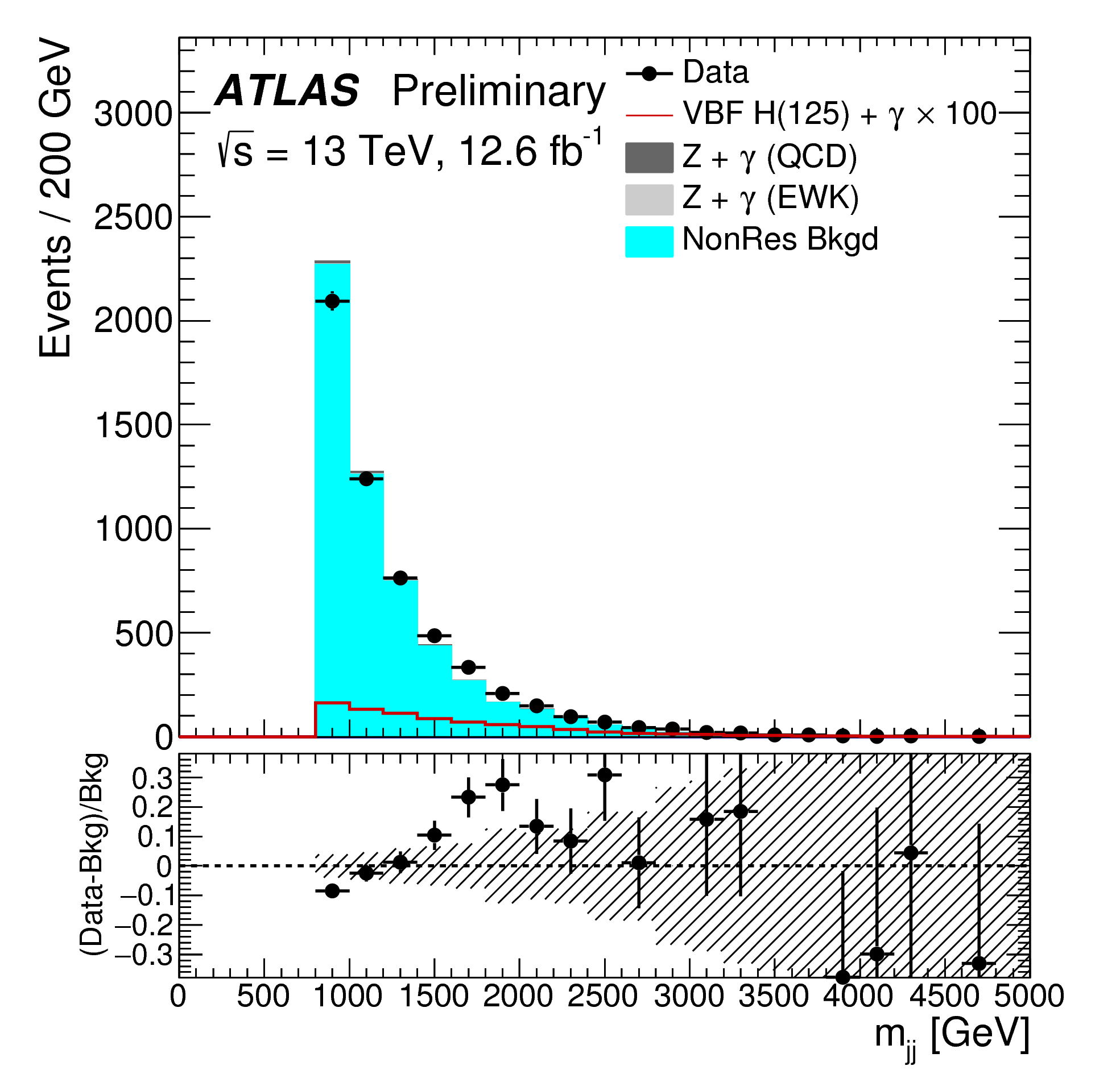}
\includegraphics[height=2in]{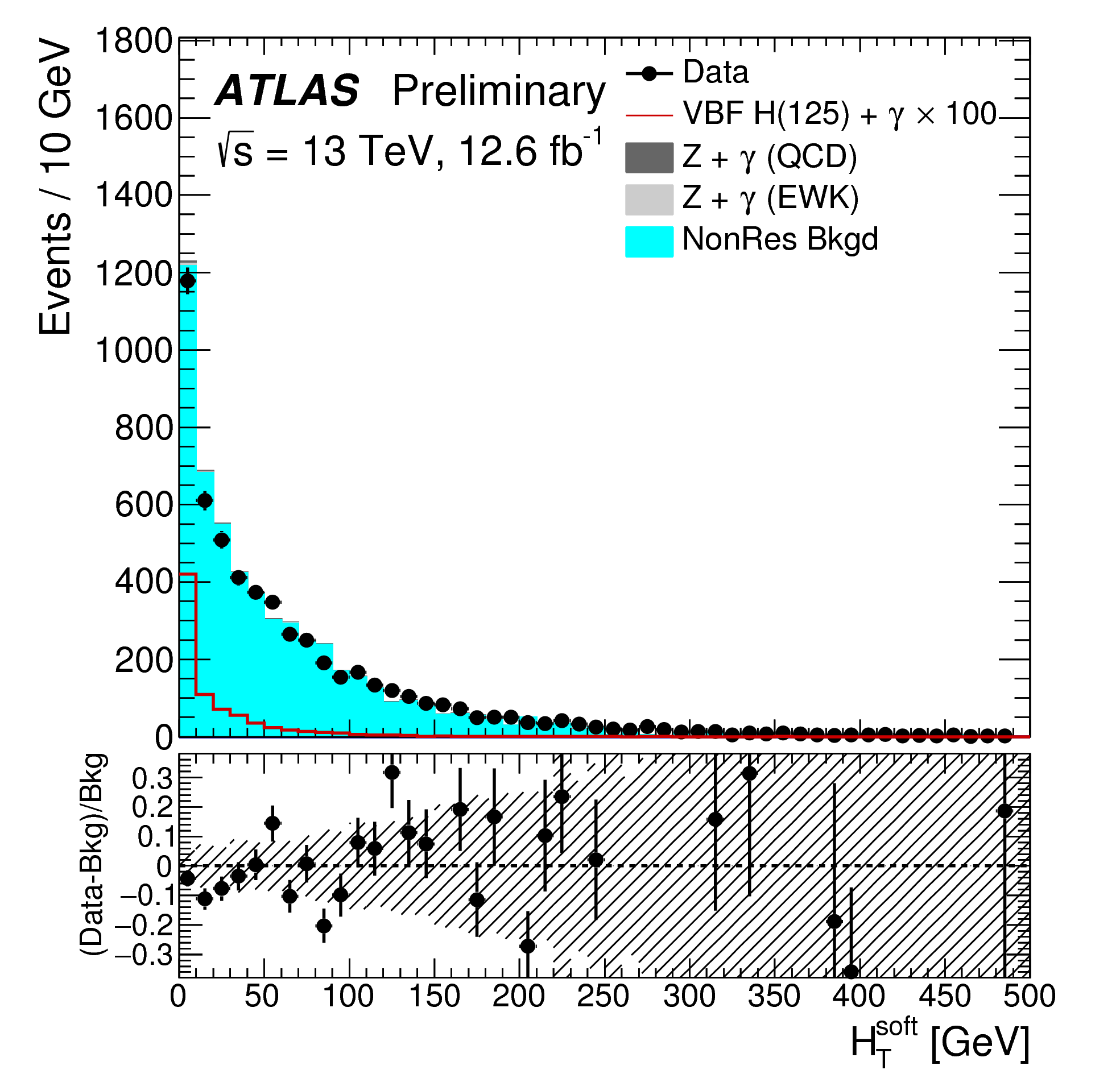}
\includegraphics[height=2in]{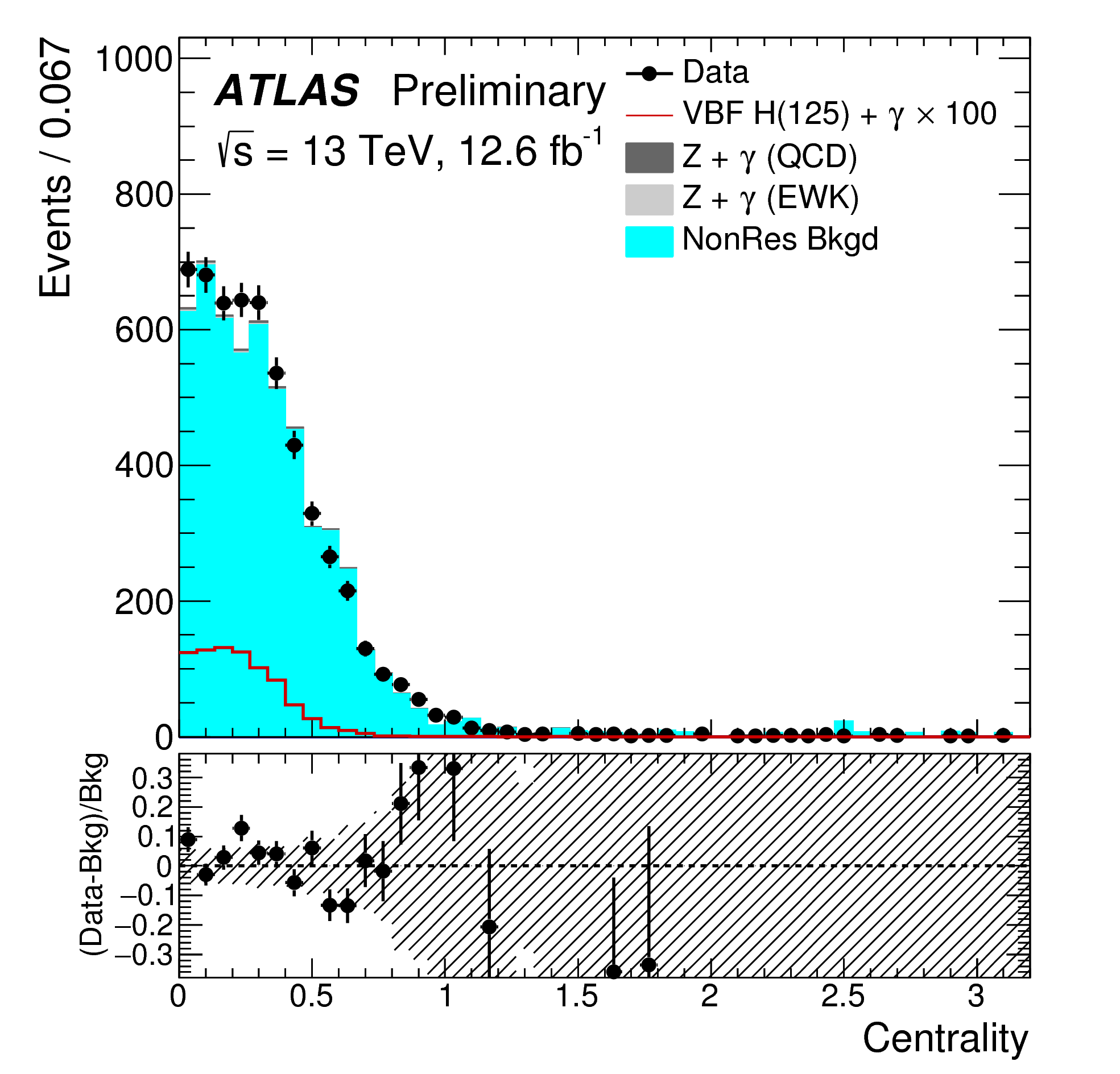}
\caption{Distributions of the input variables for the BDT training, for signal, background, and data. The di-jet mass distribution is shown in left plot. The distribution of $\mathrm{H_T}^{\mathrm{soft}}$ variable is shown in middle plot. The distribution of photon centrality is shown in the right plot.  The shaded band in the lower panel shows the statistical uncertainty on the simulated background. The signal distributions are scaled by a factor of 100. Points in the ratio outside the shown range are not displayed~\cite{ATLASVBFhbbg}.
}
\label{fig:var}
\end{figure}

  The $\mathrm{H_T}^{\mathrm{soft}}$ and the photon centrality variables have the best discriminant power besides the dijet mass $m_{jj}$ variable. The distribution of $\mathrm{H_T}^{\mathrm{soft}}$ and the photon centrality for signal, background simulation and data are shown in Fig.~\ref{fig:var}. 

\begin{figure}[htb]
\centering
\includegraphics[height=3in]{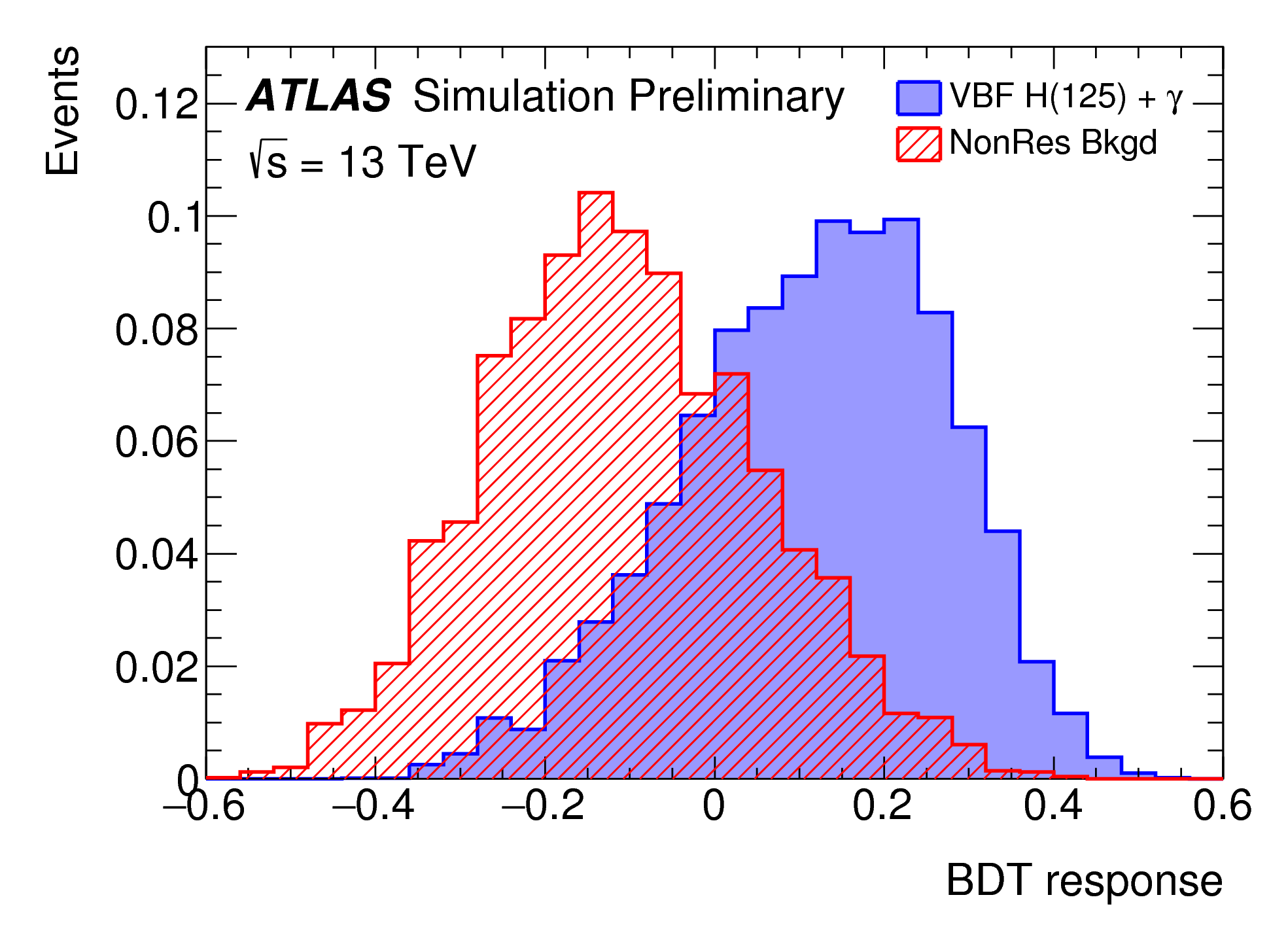}
\caption{Distribution of the BDT output response in simulated $H\gamma j j$  signal events and non-resonant $b\bar{b}\gamma j j$ background events. The distributions are normalized to unit area~\cite{ATLASVBFhbbg}.
}
\label{fig:BDT}
\end{figure}

\begin{figure}[htb]
\centering
\includegraphics[height=3in]{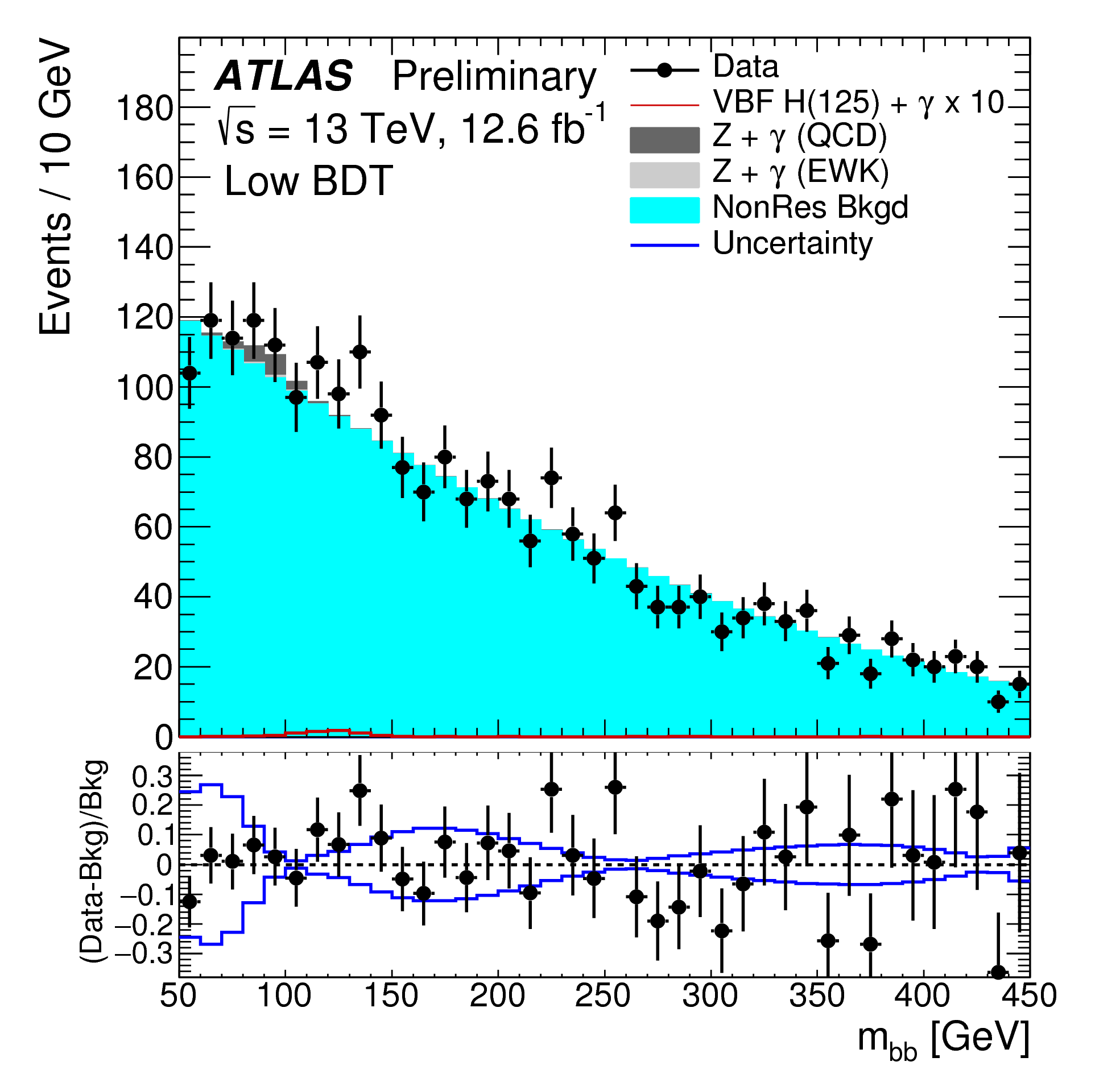}
\includegraphics[height=3in]{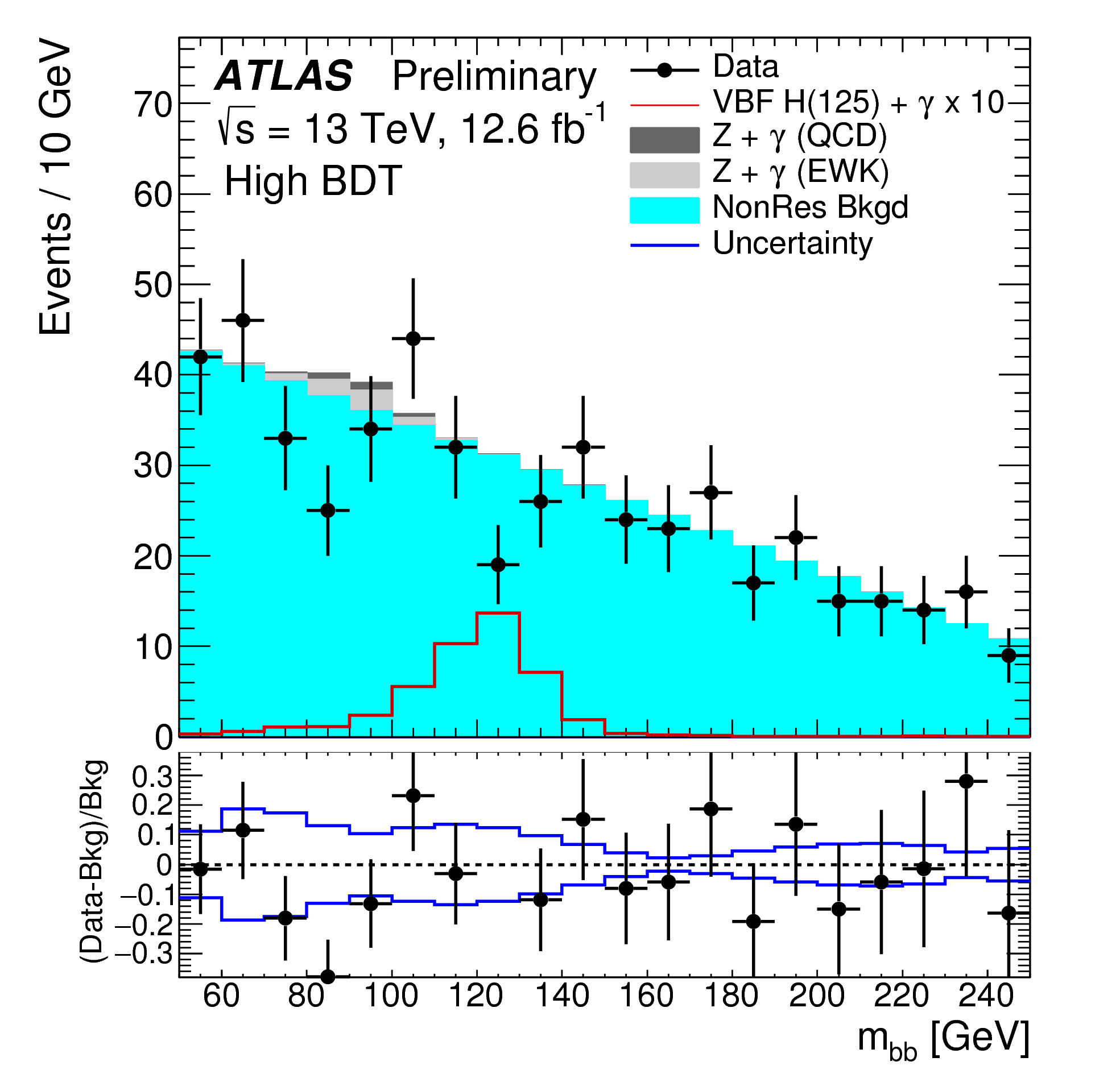}

\caption{Invariant mass distributions for each of the three BDT regions considered in the likelihood fit. Contributions are included from the Higgs boson signal, $Z+\gamma$ production through strong and electroweak processes, and non-resonant $b\bar{b}$ background. The Higgs boson signal distributions are scaled to signal strength $\mu=10$. The blue line in the lower panels shows the statistical and systematic uncertainties combined~\cite{ATLASVBFhbbg}.
}
\label{fig:mbb2}
\end{figure}

The distribution of the BDT response for the Higgs signal and the simulated backgrounds is shown in Fig.~\ref{fig:BDT}. Three different regions are defined using the BDT score: low ($-1.0$ $<\mathrm{BDT}<$ $-0.1$), medium ($-0.1$ $<\mathrm{BDT}<$ 0.1) and high ( 0.1 $<\mathrm{BDT}<$ 1.0). The invariant mass distribution of the b-jet pairs for the low and high BDT score regions are shown in Fig.~\ref{fig:mbb2}. The expected SM $H \to b\bar{b}$ distribution multiplied by a factor ten is shown as a solid line. The data agrees well with the background expectations as shown in the lower panels of the figure. The b-jet pairs invariant mass distributions of events falling in these three categories are fitted using a profiled likelihood fit, where the non-resonant background is represented by a smoothly- falling polynomial distribution and the signal with a Crystal Ball function. The fit is tested by searching for the SM $Z \to b\bar{b}$ signal produced in VBF in association with a photon. The expected significance in the $Z(\to b\bar{b})+\gamma j j$ search was 1.3 while the observed value was 0.4. The fitted signal strength was $\mu_{Z(\to b \bar{b})+\gamma j j} =0.3\pm0.8$. For the Higgs boson search,the expected 95\%CL limit was $6^{+2.3}_{-1.7}$ times the SM expectation, while the observed one was 4.0 times the SM prediction. The dominant systematic uncertainties come from the background estimation, theoretical uncertainties on the signal acceptance and cross section and jet energy scale.

\section{Conclusions}
The search for the Higgs boson decaying to b-quark pairs is essential to probe the nature of the Higgs boson and test the predictions of the Standard Model, since it can probe the Yukawa coupling to down-type quarks and can be used to constrain the total width of the Higgs.
For the $VH$ associated production, ATLAS presented an evidence for a Standard Model Higgs boson decaying into a $b\bar{b}$ pair and produced in association with a $W$ or $Z$ boson, using 36.1 $\mathrm{fb^{-1}}$  data collected by the ATLAS experiment in proton–proton collisions from Run 2 of the Large Hadron Collider. The expected significance is 3.0, while the observed one is 3.5, corresponding to a signal strength of $\mu_{VZ}$ = $1.2 ^{+0.24}_{-0.23}$(stat.)$^{+0.34}_{-0.28}$(syst.).

ATLAS also presented the first result for searching for the Higgs boson decaying to b-quark pairs in vector boson fusion process with an extra high-$p_T$ photon, using 13.2 $\mathrm{fb^{-1}}$  of 13 $\mathrm{TeV}$ $pp$ collisions. The expected 95\% CL limit of this search is $6^{+2.3}_{-1.7}$ times the SM expectation, while the observed one is 4.0.

%%  if necessary
%\Acknowledgements
%I am grateful to XYZ for fruitful discussions.

\end{document}